 \documentclass[final,5p,times,twocolumn,authoryear]{elsarticle}

\usepackage{amssymb}
\usepackage{lipsum}
\usepackage{amsmath}
\usepackage{multirow}
\usepackage{array}
\usepackage{graphicx}
\usepackage{rotating}
\usepackage{url} 
\usepackage{hyperref}
\usepackage{xcolor}
\usepackage{adjustbox}

\journal{High Energy Astrophysics}

\begin{document}

\begin{frontmatter}



\title{Microquasars to AGNs: An uniform Jet variability}




\author[first]{Ajay Sharma}
\affiliation[first]{{S N Bose National Centre for Basic Sciences,},
            addressline={Block JD, Salt Lake}, 
            city={Varanasi},
           postcode={700106}, 
            state={West Bengal},
            country={India}}

\author[second]{Raj Prince}
\affiliation[second]{{Department of Physics, Institute of Science, Banaras Hindu University,},
            city={Varanasi},
           postcode={221005}, 
            state={Uttar Pradesh},
            country={India}}

\author[fourth]{Debanjan Bose}
\affiliation[fourth]{{Department of Physics, Central University of Kashmir,},
            addressline={Ganderbal}, 
           postcode={191131}, 
            state={Kashmir},
            country={India}}            

\begin{abstract}
    The long-term variability study over a range of black hole (BH) mass systems from the microquasars of stellar-mass black holes to the Active Galactic Nuclei (AGNs) of supermassive black holes, in $\gamma$-rays offers new insights into the physics of relativistic jets. In this work, we investigate the $\gamma$-ray variability of 11 AGNs--including 7 blazars, 2 unclassified blazar candidates (BCUs), 1 radio galaxy (RG), and 1 narrow-line Seyfert 1 galaxy (NLS1) as well as 2 microquasars. We apply a stochastic process known as the Damped Random Walk (DRW) to model the $\sim$15 years of Fermi-LAT light curves. The characteristic timescales observed for AGNs are comparable to those in the accretion disc. Interestingly, the timescales observed in the jet emission of microquasars are similar to those of AGNs, suggesting uniform jet properties across the black hole masses. 
    The observed rest-frame timescales of AGNs overlap with both thermal and non-thermal timescales associated with the jet and accretion disk, respectively, suggesting a scaled relationship between $\tau_{DRW}^{rest}$ and black hole mass ($\rm{M_{BH}}$). While the timescales observed for microquasars deviate significantly from this relationship, nonetheless exhibit a scaled $\tau_{DRW}^{rest}-\rm{M_{BH}}$ relationship using $\gamma$-rays specifically. These findings offer new insights into the origin of jets and the processes driving the emission within them. Additionally, this study hints at a new perspective that the relativistic jets' properties or their production mechanisms may be independent of the black hole mass.
\end{abstract}




\end{frontmatter}




\section{Introduction}
\label{introduction}

Since the launch of Fermi Large Area Telescope (Fermi-LAT) it has become a premier instrument for studying both galactic and extragalactic sources in the high-energy regime, particularly at energies of 100 MeV and above. Observations have shown that $\gamma$-ray emissions from Active Galactic Nuclei (AGNs) dominate the extragalactic $\gamma$-ray sky. AGNs are among the most luminous objects in the universe, powered by supermassive black holes (SMBHs) at their cores. The accretion of matter onto these SMBHs is the primary energy source driving the immense luminosity of AGNs. Based on their observational properties—such as bolometric luminosity, flux variability, and broadband spectral energy distribution (SED)—AGNs are categorized into various classes.\par

A particularly extreme class of AGNs, known as blazars, exhibits strong Doppler-boosted non-thermal emission and rapid, high-amplitude variability across the entire electromagnetic spectrum, with timescales ranging from minutes to years. Blazars are unique in that their relativistic jets are closely aligned with the observer's line of sight (within approximately 5°) \cite{antonucci1993unified, urry1995unified}. Blazars are further subdivided into two categories: BL Lac objects (BL Lacs) and Flat Spectrum Radio Quasars (FSRQs), distinguished primarily by the strength of their optical emission lines \cite{fossati1998unifying, ghisellini2017fermi, ajello2020fourth}. \par

AGN variability has been detected across the entire electromagnetic spectrum, including in jetted AGNs with relatively large inclination angles—referred to as misaligned AGNs or radio galaxies. These objects also exhibit strong and persistent flux variability in $\gamma$-rays \cite{magic2018broad}. Numerous studies have been conducted to explore the underlying physical processes that characterize the variability in these sources \cite{abdalla2017characterizing, yan2018statistical, rieger2019gamma, bhatta2020nature}. For instance, the Narrow-Line Seyfert 1 (NLS1) galaxy PMN J0948+0022 has been observed in gamma rays at energies above 100 MeV \cite{foschini2012radio}. This source has also been extensively investigated using multiwavelength observations \cite{d2014multiwavelength}. 
Lower-mass black hole systems, such as microquasars and X-ray binaries, have also been studied through multiwavelength observations \cite{acciari2009multiwavelength}. \par
$\gamma$-ray variability studies have become a crucial tool for understanding the physical processes within relativistic jets from AGNs and even microquasars. Flux variability is observed across a wide range of timescales. One effective method for quantifying this variability is through the power spectral density (PSD), which measures the amplitude of variations in a time series as a function of frequency or timescale. The PSD is particularly useful for characterizing the complex nature of emissions, which may be driven by an underlying stochastic process. In some cases, emissions exhibit an oscillatory pattern known as quasi-periodic oscillations (QPOs), which have been detected in Fermi-LAT data \cite{ackermann2015multiwavelength, sandrinelli2016quasi, zhou201834, penil2020systematic, banerjee2023detection, prince2023quasi}. However, the reliability of these QPOs remains questionable, and the exact mechanisms behind their production are still not well understood. \par

Another approach involves to understand AGN variability by modeling it with stochastic process, which has been extensively used to describe the optical variability of AGNs \cite{kelly2009variations, macleod2010modeling, zu2013quasar, rakshit2017optical, li2018new, zhang2018broadband, lu2019supermassive, burke2021characteristic, zhang2023gaussian}. This method has proven particularly useful for characterizing variability across multiple wavelengths in AGN emissions. The stochastic process is typically represented by the Damped Random Walk (DRW) model, also known as the Ornstein-Uhlenbeck (OU) process, which is characterized by a break-like feature in its PSD. Generally, the PSDs of jets and accretion discs are described by a bending power law (BPL). At frequencies above the break frequency (which corresponds to a characteristic timescale), the PSD follows a power law with a slope of $\sim$2. Below the break frequency, the PSD is represented by white noise. This stochastic model effectively captures the long-term variability of AGN accretion discs.\par

In recent years, the DRW model has been widely applied not only in the optical time domain but also to submillimeter \cite{chen2023testing}, X-ray \cite{zhang2024discovering}, and $\gamma$-ray \cite{sobolewska2014stochastic, goyal2018stochastic, ryan2019characteristic, tarnopolski2020comprehensive, covino2020looking, yang2021gaussian, zhang2022characterizing, sharma2024probing} variability in AGNs and even microquasars. These findings suggest that non-thermal timescales from jets are somehow connected to the thermal timescales of accretion discs.\par

The Fermi-LAT observatory has been successfully operating and observing the entire sky for over 15 yrs, offering an exceptional opportunity to study long-term gamma-ray variability in AGNs and microquasars. In this work, we apply the \texttt{celerite}\footnote{\url{https://celerite.readthedocs.io/en/stable/}} model \cite{foreman2017fast} to the nearly 15-yr Fermi-LAT light curves of 13 sources, including 7 blazars, 2 blazar candidates of uncertain type (BCUs), 1 radio galaxy, 1 Narrow-Line Seyfert 1 (NLS1) galaxy, and 2 microquasars. Our goal is to investigate the gamma-ray variability of these sources, explore the underlying emission processes, and establish connections between observational data and theoretical models.

The structure of this paper is as follows: Section \ref{sec:data} provides a brief overview of the Fermi-LAT data processing methods. Section \ref{sec:model} introduces the DRW model. In Section \ref{sec:result}, we summarize the DRW modeling results for all sources in our sample. In Section \ref{sec:discussion}, we discuss the observed results in detail.

\section{Sample and Fermi-LAT Data Analysis}\label{sec:data}

The sample in this investigation consists of 13 sources, including 12 Active Galactic Nuclei (AGNs). Among these AGNs, there are 7 Blazars, 1 Radio galaxy, 1 Narrow-line Seyfert 1 galaxies, and 2 Unclassified blazar candidates. Each of these sources demonstrates significant variability in $\gamma$-ray emissions, with detailed information available in Table \ref{tab-1}.\par
The Fermi-LAT observations \cite{abdo2010spectral} for all sources in the sample were collected over a span of $\sim$15 yr (from MJD 54675 to 60460), within the energy range of 0.1-300 GeV. The analysis of the gamma-ray data was conducted using \textit{Fermi} Science Tools, \texttt{FERMITOOLS} package\footnote{\url{https://fermi.gsfc.nasa.gov/ssc/data/analysis/documentation/}}. Events from the SOURCE class (evclass=128, evtype=3) were selected within a region of interest (ROI) of 15$^\circ$ for each source using the \texttt{gtselect} tool. To avoid contamination from the Earth's limb, a maximum zenith angle of 90$^\circ$ was applied. To ensure data quality and good time intervals (GTIs), the standard criteria with the recommended filter expression \texttt{$\text{(DATA\_QUAL > 0) \&\& (LAT\_CONFIG == 1)}$} were used. The instrument response function "P8R3\_SOURCE\_V3" was utilized during data processing. The galactic diffuse emission model \texttt{gll\_iem\_v07.fits}\footnote{\label{fermi} \url{https://fermi.gsfc.nasa.gov/ssc/data/access/lat/BackgroundModels.html}} and extra-galactic diffuse emission model \texttt{iso\_P8R3\_SOURCE\_V3\_v1.txt}\footref{fermi} were employed. In the likelihood analysis, the unbinned likelihood approach\footnote{\url{https://fermi.gsfc.nasa.gov/ssc/data/analysis/scitools/likelihood_tutorial.html}} was performed using the \texttt{GTLIKE} tool \citealt{cash1979parameter, mattox1996likelihood}, which provided the significance of each source within the ROI, including the source of interest, in the form of the Test Statistic (TS), which is defined as TS = -2ln$\left(\frac{L_{max,0}}{L_{max,1}}\right)$, where $L_{max,0}$ and $L_{max,1}$ are the maximum likelihood value for a model without an additional source and the maximum likelihood value for a model with the additional source at a specified location, respectively.\par
The light curves for sources are generated with a TS greater than 9 using the unbinned likelihood approach, ensuring reliable outcomes with a high signal-to-noise ratio. During the light curve generation, parameters for sources located more than 15$^\circ$ from the center of the Region of Interest (ROI) were fixed, while those within 10$^\circ$ were allowed to vary freely. The \texttt{FERMIPY}\footnote{\url{https://fermipy.readthedocs.io/en/latest/}} software package was utilized for this analysis.

\section{Stochastic Process}\label{sec:model}

The general consensus is that the variability observed in AGNs is inherently stochastic. The variability in AGNs can be explored through \texttt{Continuous Time Autoregressive Moving Average} [CARMA(p, q)] processes \citep{kelly2014flexible}, which are defined as the solutions to the following stochastic differential equation:

\begin{equation}
\begin{split}
\frac{d^p y(t)}{dt^p} + \alpha_{p-1}\frac{d^{p-1}y(t)}{dt^{p-1}}+...+\alpha_0 y(t) =\\
\beta_q \frac{d^q \epsilon(t)}{dt^q}+\beta_{q-1}\frac{d^{q-1}\epsilon(t)}{dt^{q-1}}+...+\beta_0 \epsilon(t),
\end{split}
\end{equation}

where, the time series is given as y(t), $\epsilon$(t) represents a continuous time white-noise process, $\alpha^*$ and $\beta^*$ are the coefficients of AR and MA models, 
respectively. Parameters p and q are the order of AR and MA models, respectively.\par
When the model's parameters are set to p=1 and q=0, it corresponds to a Continuous Auto-Regressive model CAR(1), also known as the Ornstein–Uhlenbeck process. In astronomical literature, this CAR(1) model is frequently referred to as a DRW (Damped Random Walk) process and is governed by the following stochastic differential equation \citep{kelly2009variations, ruan2012characterizing, moreno2019stochastic, burke2021characteristic, zhang2022characterizing, zhang2023gaussian, sharma2024probing, zhang2024discovering}.

\begin{equation}
    \left[ \frac{d}{dt} + \frac{1}{\tau_{DRW}} \right] y(t) = \sigma_{DRW} \epsilon(t)
\end{equation}

where $\tau_{DRW}$ is the characteristic damping time-scale of the DRW process and
$\sigma_{DRW}$ is representing the amplitude of random perturbations. The covariance function of the DRW model is defined as
\begin{equation}
    k(t_{nm}) = a.exp(-t_{nm}c),
\end{equation}
where $t_{nm} = | t_n -t_m|$ represents the time lag between measurements m and n, with $a = 2 \sigma_{DRW}^2$ and $c = \frac{1}{\tau_{DRW}}$. The power spectral density (PSD) is expressed as:
\begin{equation}
    S(\omega) = \sqrt{\frac{2}{\pi}} \frac{a}{c} \frac{1}{1 + (\frac{\omega}{c})^2}
\end{equation}

The PSD of DRW is a Broken Power Law (BPL) form, where the broken frequency $f_b$ corresponds to the characteristic damping timescale $\tau_{DRW} = \frac{1}{2\pi f_b}$.\par

To estimate the best-fit parameters and their associated uncertainties of the DRW model for each source in our sample, we used the Markov chain Monte Carlo (MCMC) algorithm provided by the \textsc{emcee}\footnote{\url{https://emcee.readthedocs.io/en/stable/}} package \cite{foreman2013emcee}. For the \textsc{celerite} modeling, we executed MCMC with 32 parallel chains, each running 10,000 steps for burn-in and 20,000 steps for generating the parameter distributions. In the \textsc{emcee} modeling, we determined the Maximum a posteriori (MAP) parameters using the nonlinear optimizer \textsc{L-BFGS-B}\footnote{\url{https://docs.scipy.org/doc/scipy/reference/optimize.minimize-lbfgsb.html}}, implemented by the \textsc{Scipy} project.

\begin{figure*}
    \centering
    \includegraphics[width=0.85\textwidth]{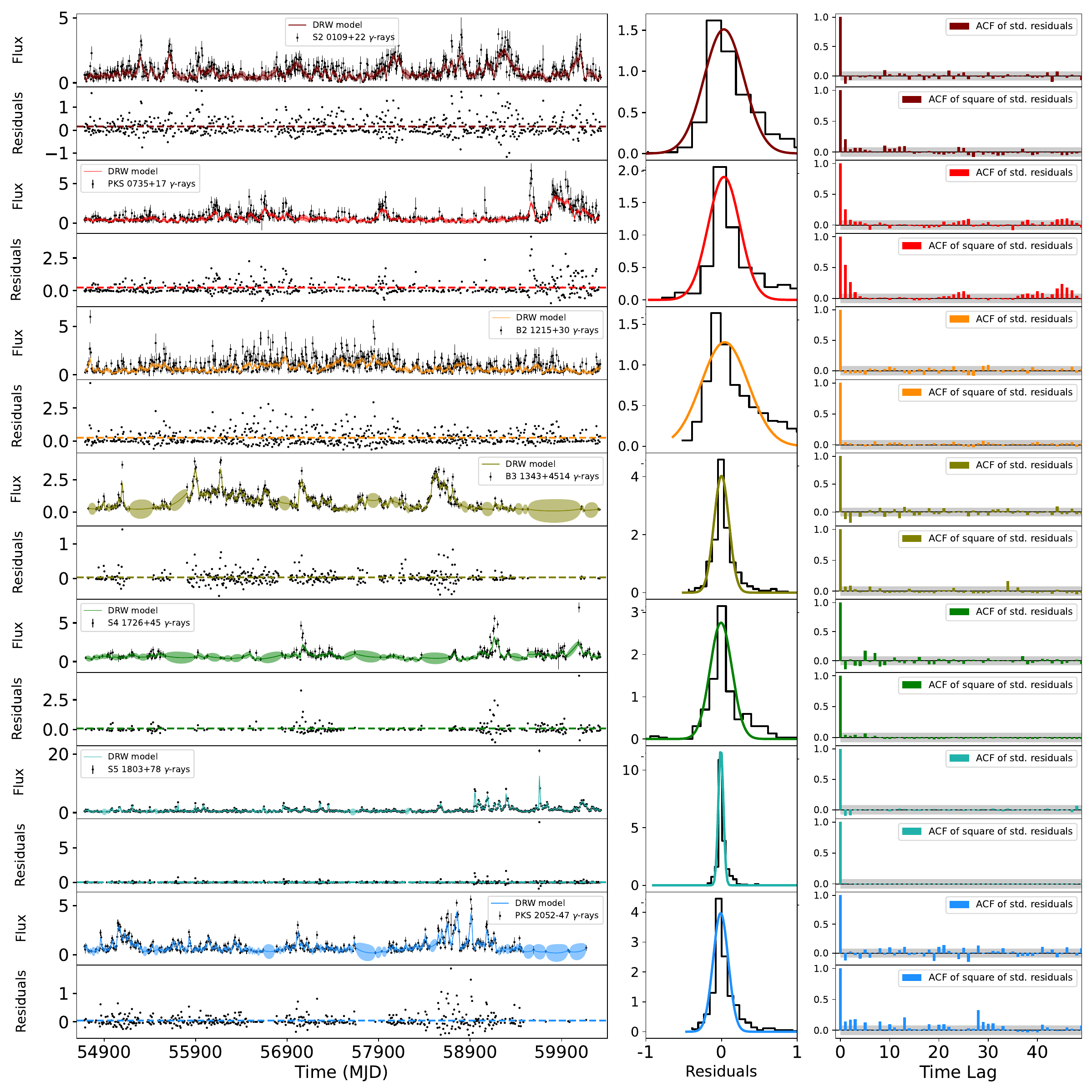}
    \caption{Results of the DRW model fitting for all blazars in our sample, Table \ref{tab-1}. Each source results are depicted in three columns: the leftmost column shows the observed $\gamma$-ray flux (black points) and the DRW-modeled light curve (in assigned colors). The bottom panel presents the residuals (black dots) with their mean in the corresponding color. The middle column displays the residuals' distribution (black histogram) alongside the best-fit normal distribution (in corresponding color). The rightmost column features the autocorrelation functions (ACFs) of the residuals and squared residuals (in assigned colors) with the 95$\%$ confidence interval for white noise (in grey). The results from modeling all blazars with the DRW model are displayed in distinct colors for enhanced visual clarity.}
    \label{Fig-1}    
\end{figure*}

\begin{figure*}
    \centering
    \includegraphics[width=0.85\textwidth]{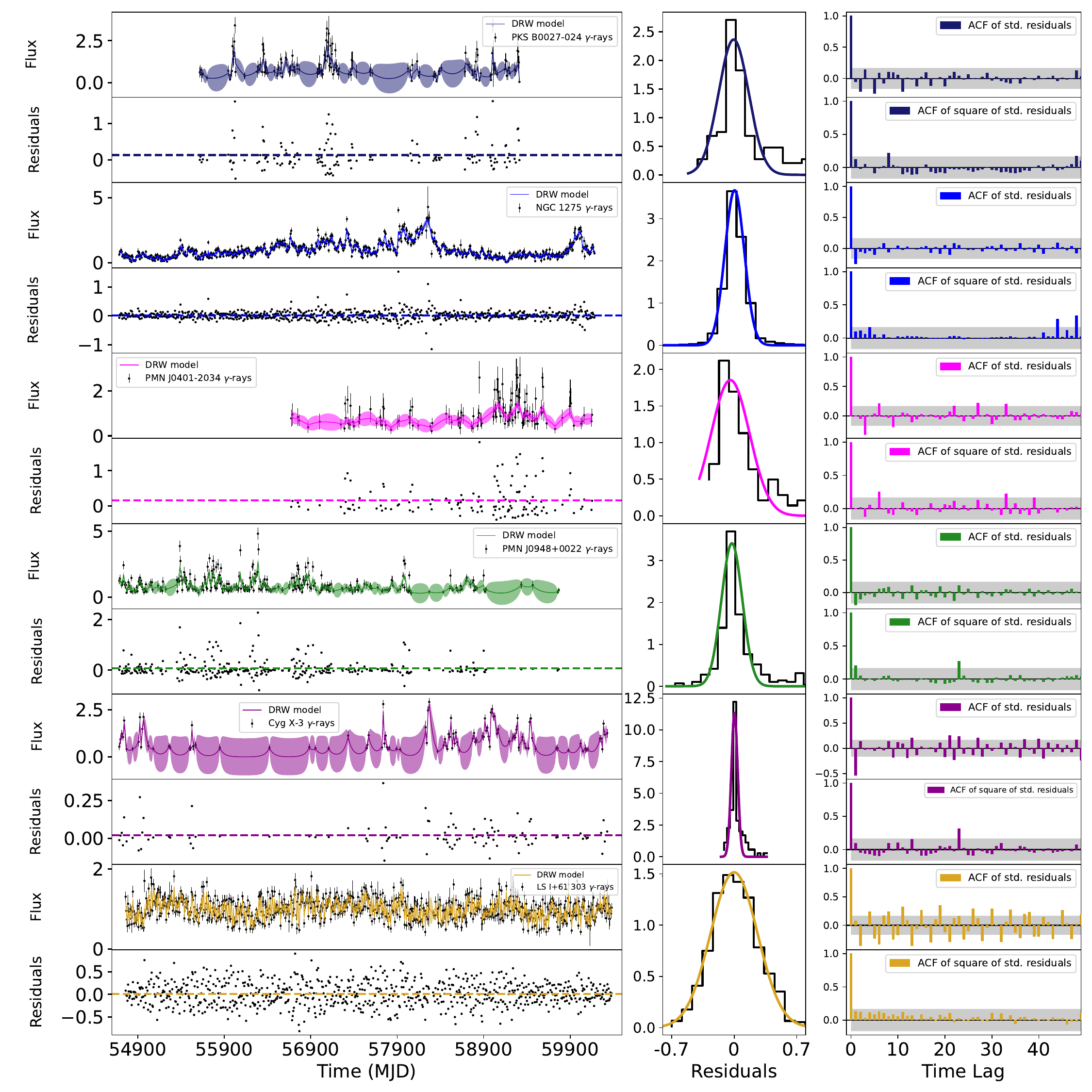}
    \caption{Results of DRW model fitting for all non-blazar sources in our sample are shown in Table \ref{tab-1}. The results are displayed in different colors as those in Figure \ref{Fig-1}.}
    \label{fig-2}
\end{figure*}

\begin{figure*}
    \centering
    \adjustbox{frame}{\includegraphics[width=0.24\textwidth]{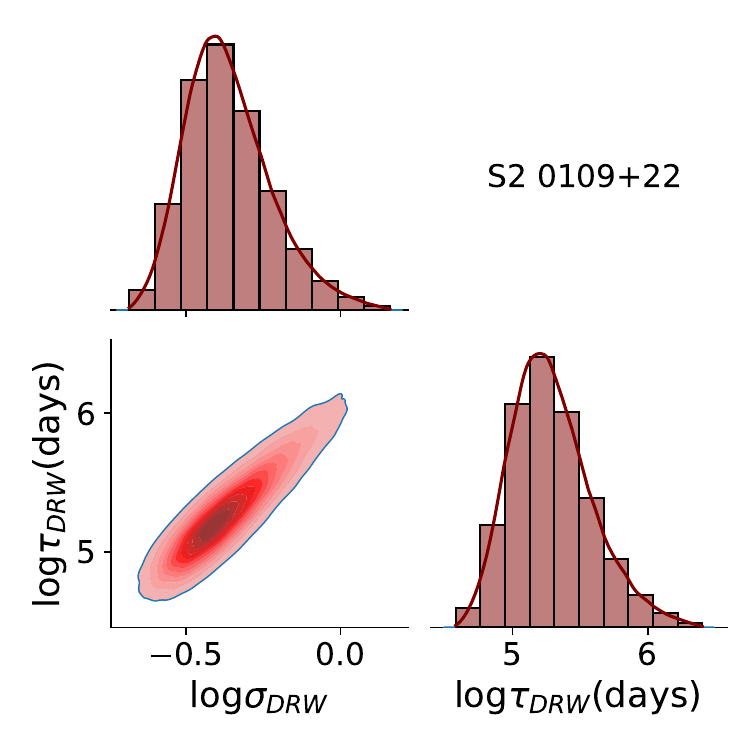}}\hspace{1pt}
    \adjustbox{frame}{\includegraphics[width=0.24\textwidth]{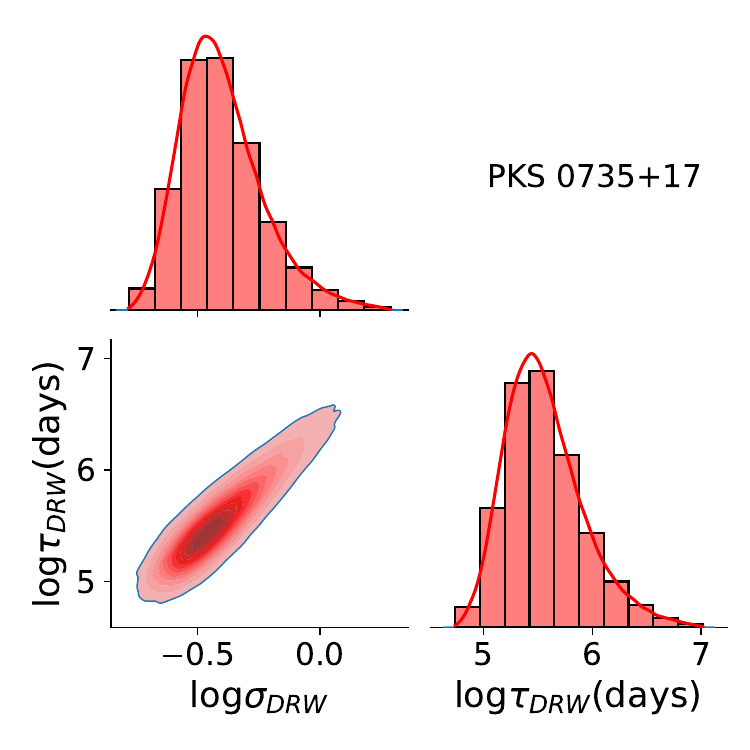}}\hspace{1pt}
    \adjustbox{frame}{\includegraphics[width=0.24\textwidth]{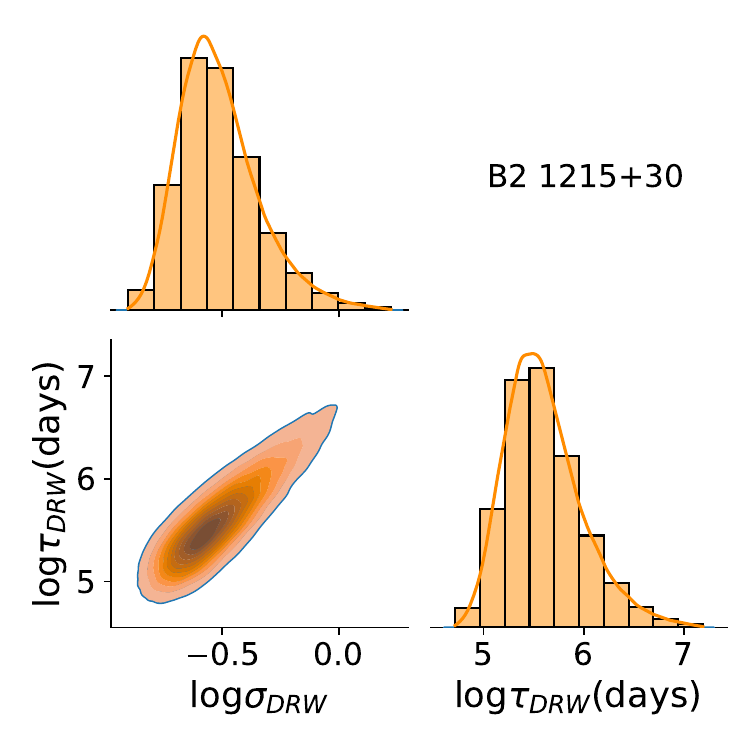}}\hspace{1pt}
    \adjustbox{frame}{\includegraphics[width=0.24\textwidth]{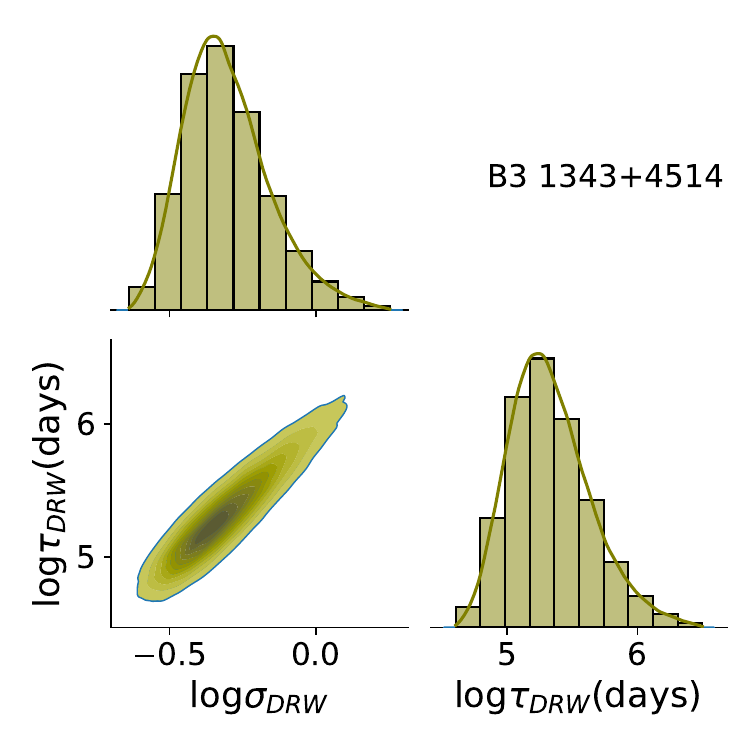}}\vspace{1pt}
    \adjustbox{frame}{\includegraphics[width=0.24\textwidth]{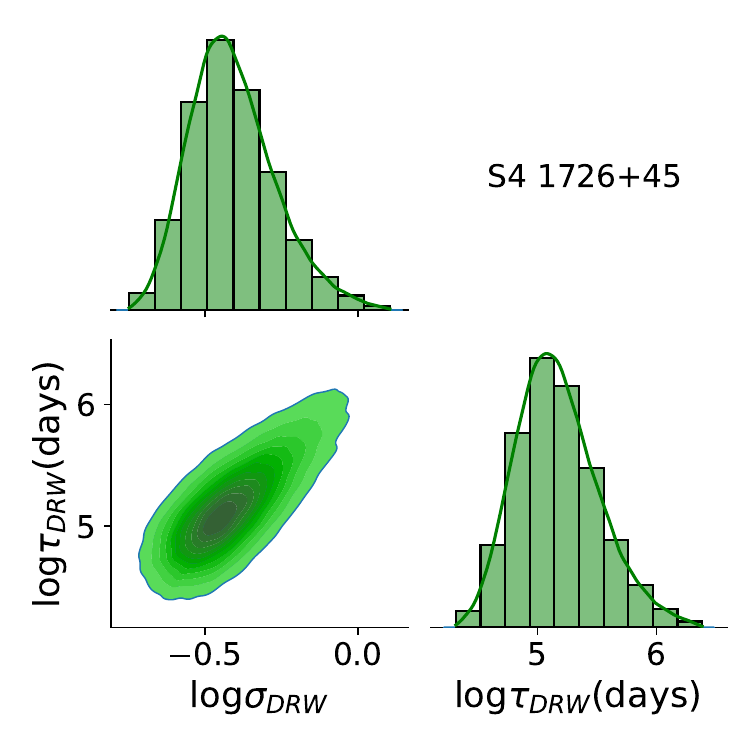}}\hspace{1pt}
    \adjustbox{frame}{\includegraphics[width=0.24\textwidth]{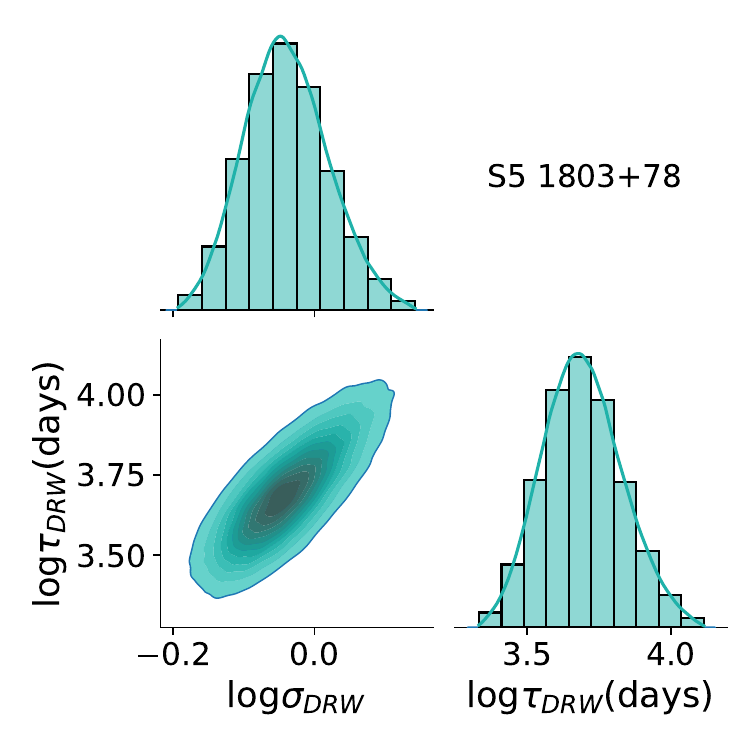}}\hspace{1pt}
    \adjustbox{frame}{\includegraphics[width=0.24\textwidth]{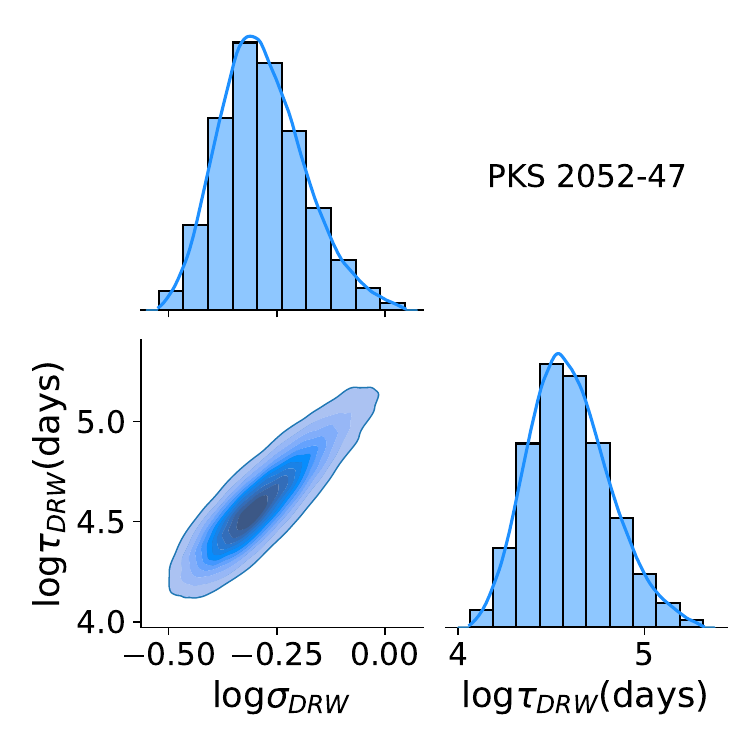}}\vspace{1pt}

    \adjustbox{frame}{\includegraphics[width=0.24\textwidth]{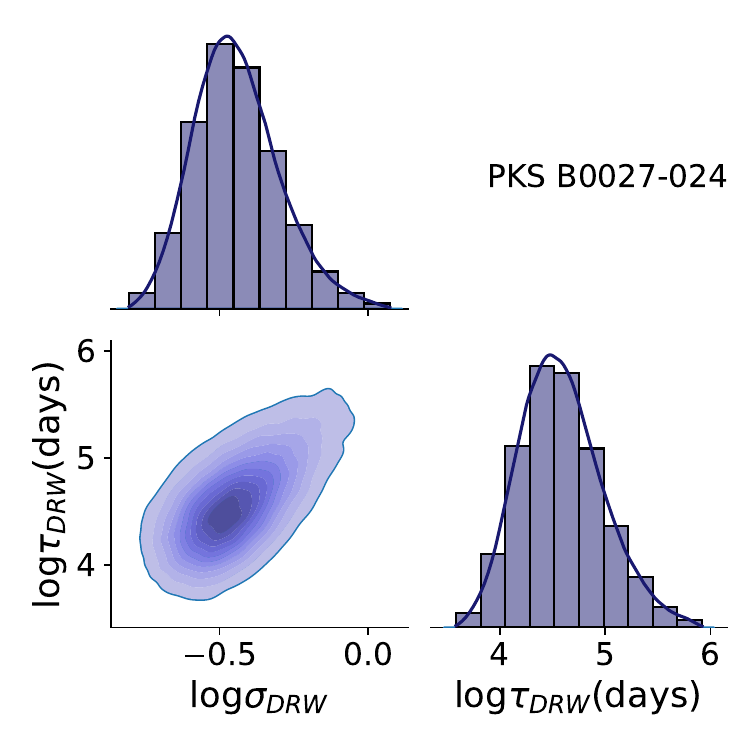}}\hspace{1pt}
    \adjustbox{frame}{\includegraphics[width=0.24\textwidth]{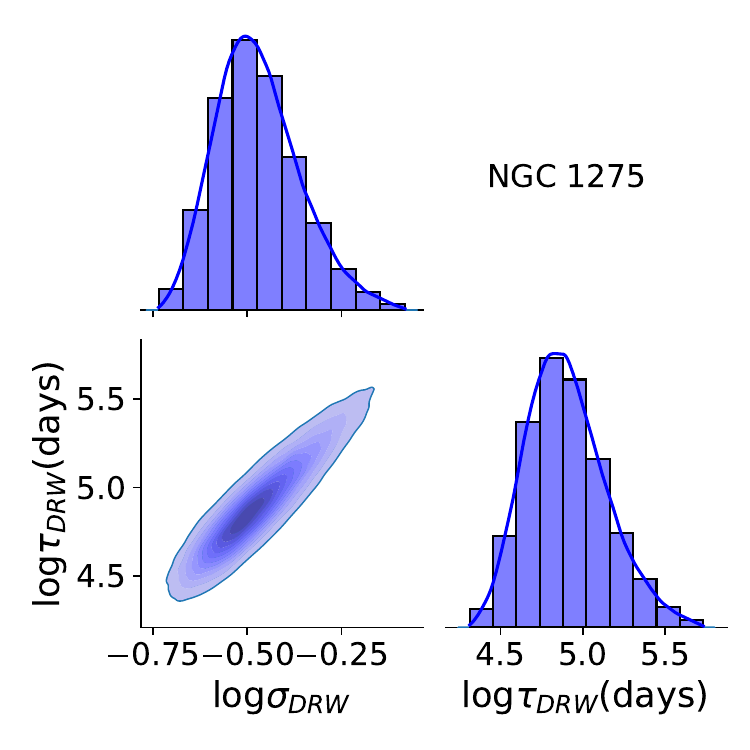}}\hspace{1pt}
    \adjustbox{frame}{\includegraphics[width=0.24\textwidth]{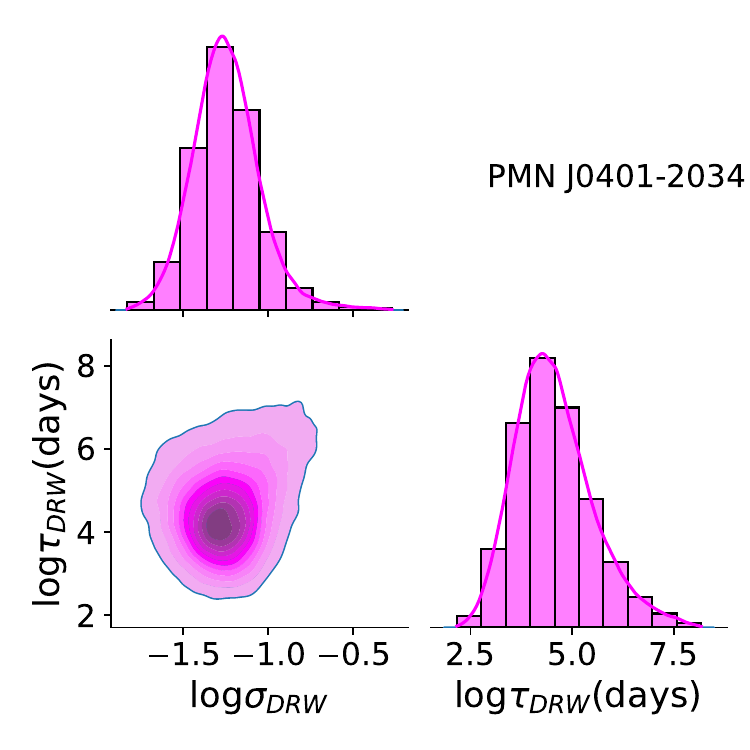}}\hspace{1pt}
    \adjustbox{frame}{\includegraphics[width=0.24\textwidth]{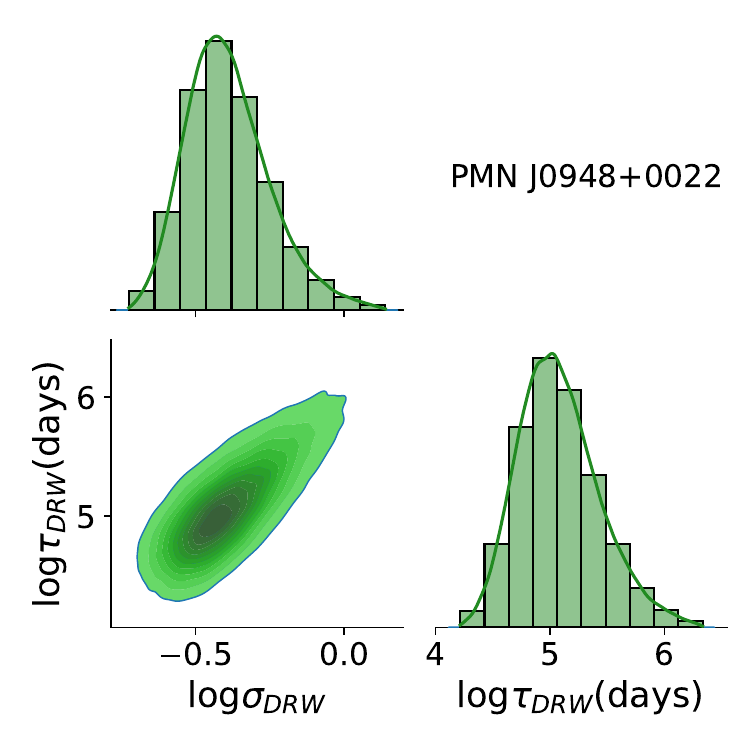}}\vspace{1pt}
    \adjustbox{frame}{\includegraphics[width=0.24\textwidth]{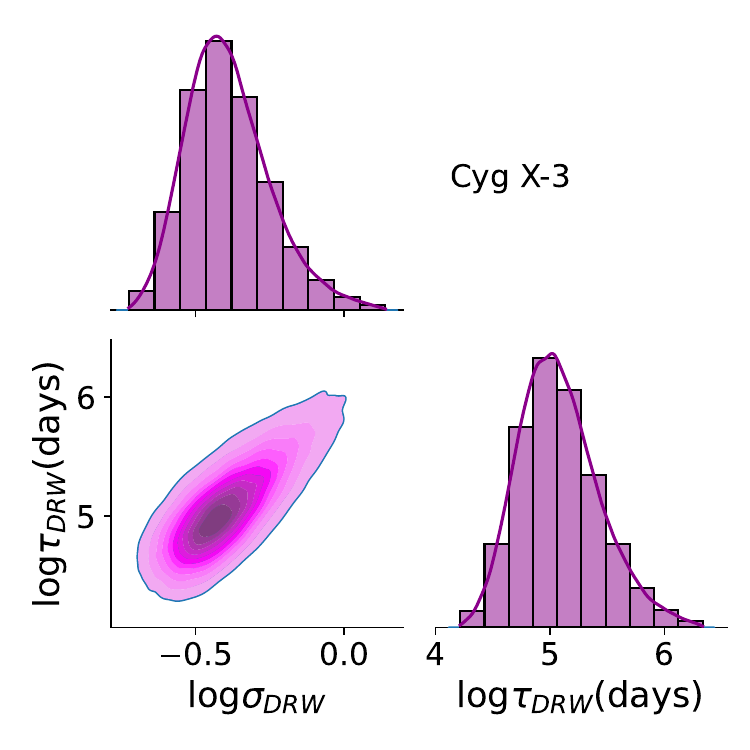}}\hspace{1pt}
    \adjustbox{frame}{\includegraphics[width=0.24\textwidth]{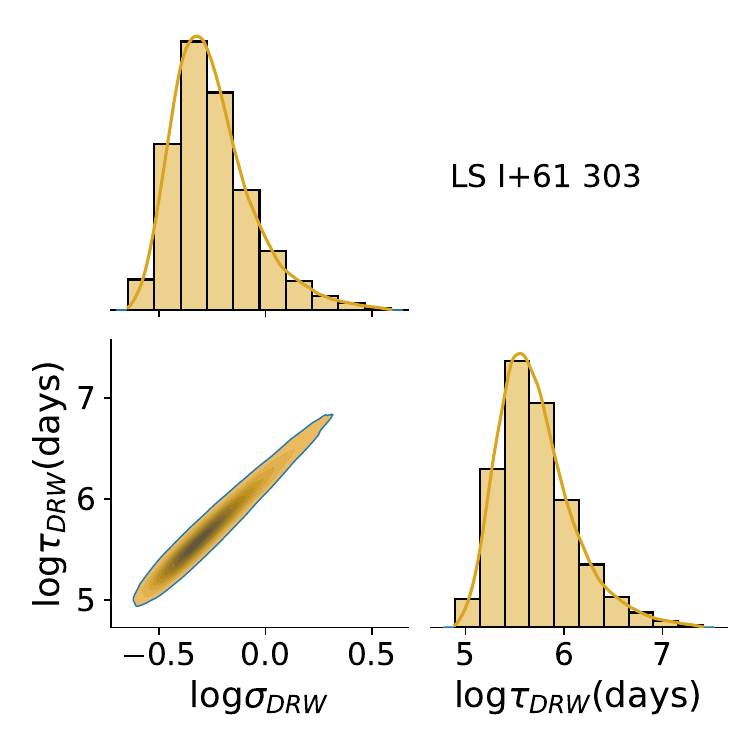}}
    \caption{Posterior probability densities of the DRW model parameters for all sources in our sample, represented with the corresponding colors.}
    \label{Fig-3}    
\end{figure*}

\begin{figure*}
    \centering
    \includegraphics[width=0.24\textwidth]{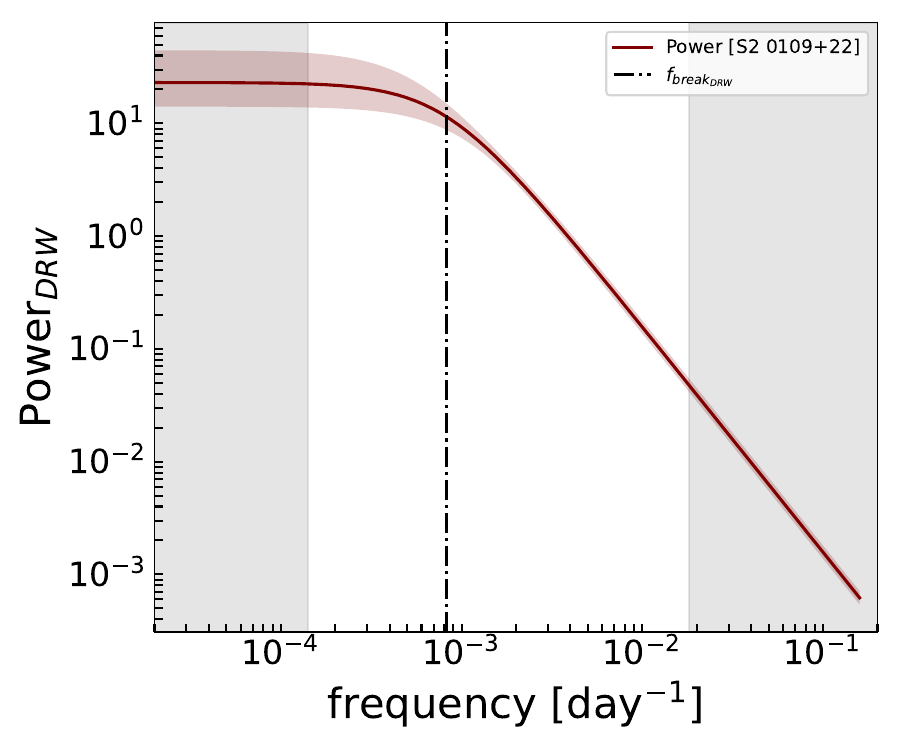}\hspace{1pt}
    \includegraphics[width=0.24\textwidth]{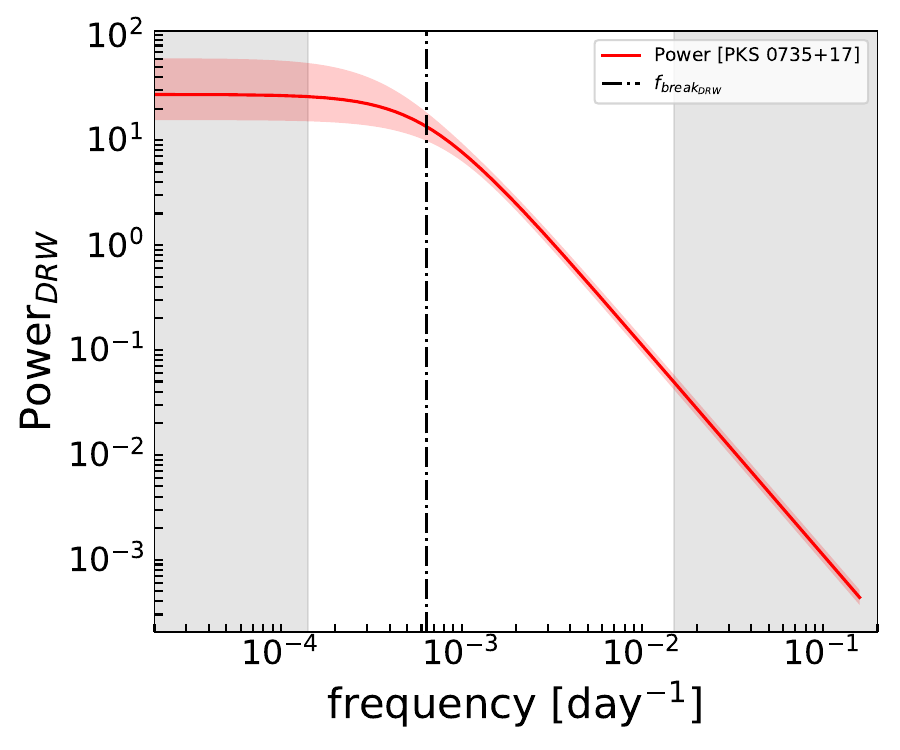}\hspace{1pt}
    \includegraphics[width=0.24\textwidth]{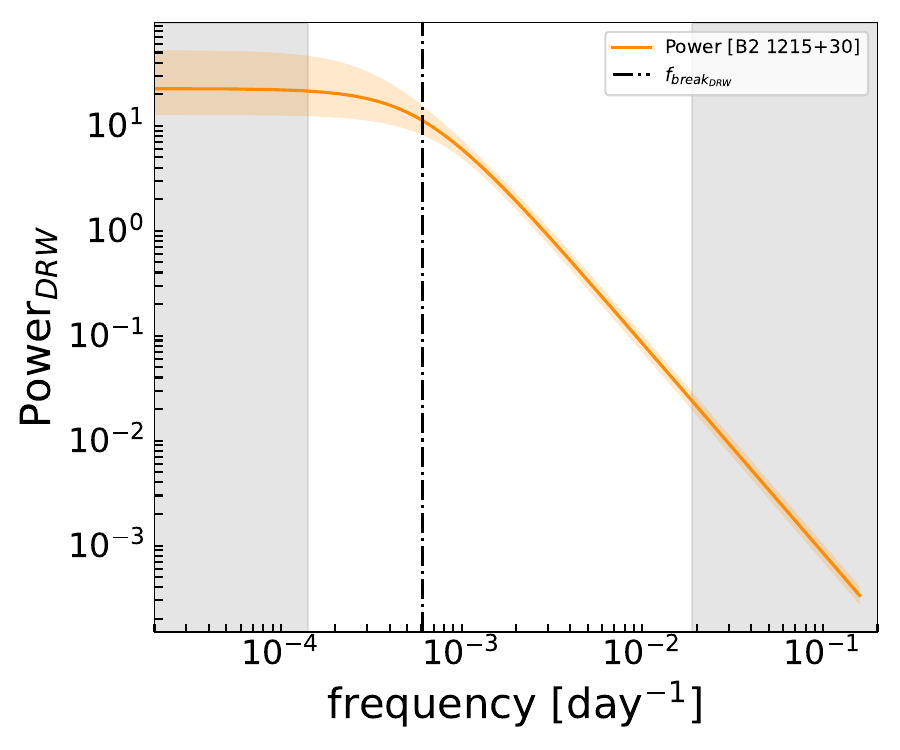}\hspace{1pt}
    \includegraphics[width=0.24\textwidth]{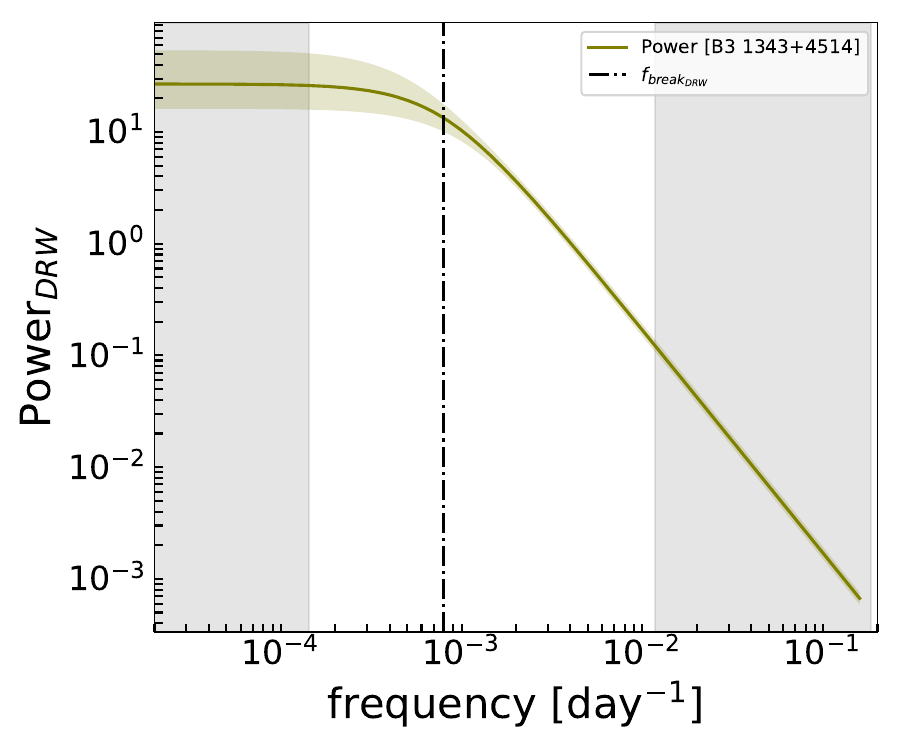}\vspace{1pt}
    \includegraphics[width=0.24\textwidth]{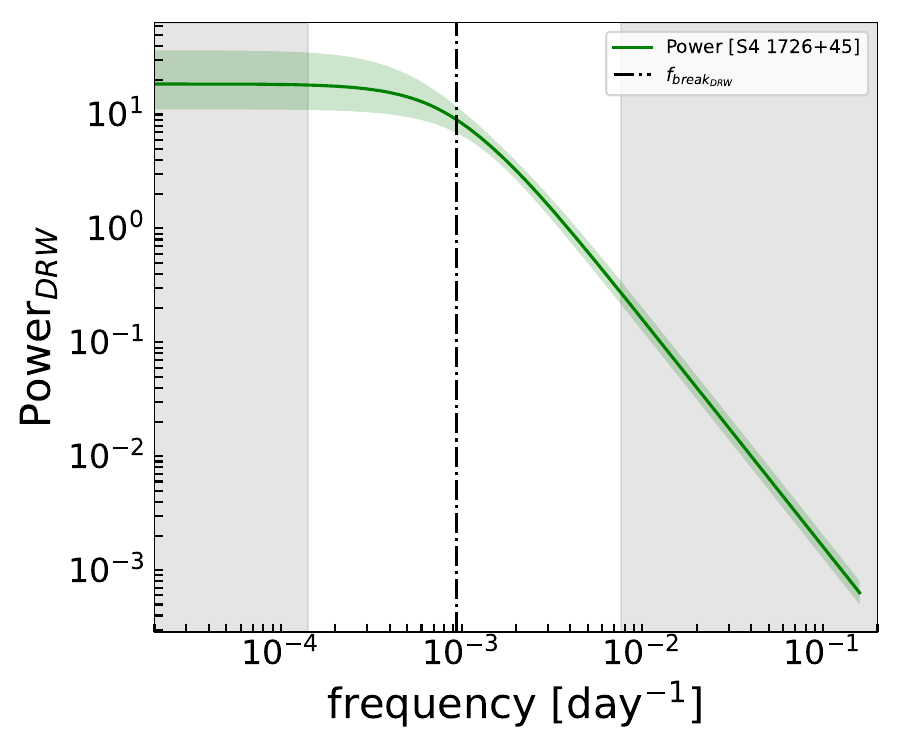}\hspace{1pt}
    \includegraphics[width=0.24\textwidth]{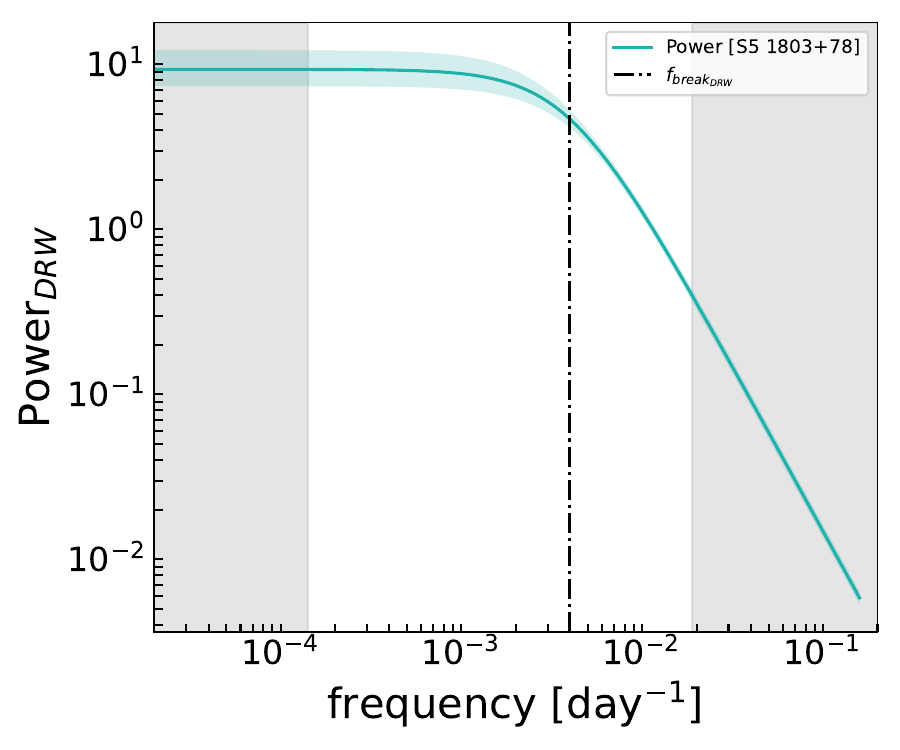}\hspace{1pt}
    \includegraphics[width=0.24\textwidth]{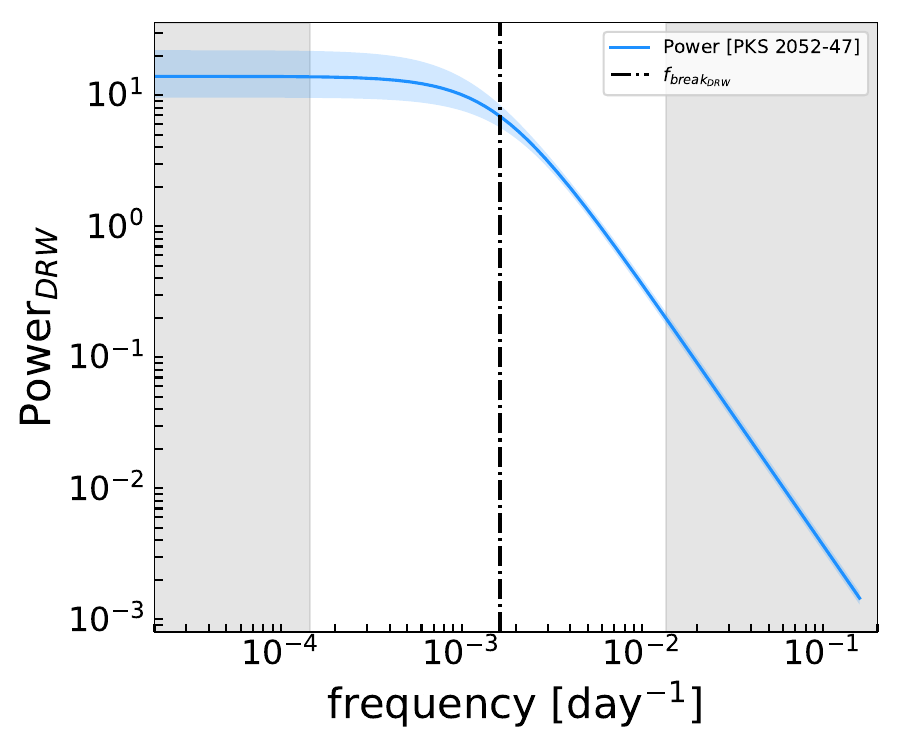}\vspace{1pt}
    
    \includegraphics[width=0.24\textwidth]{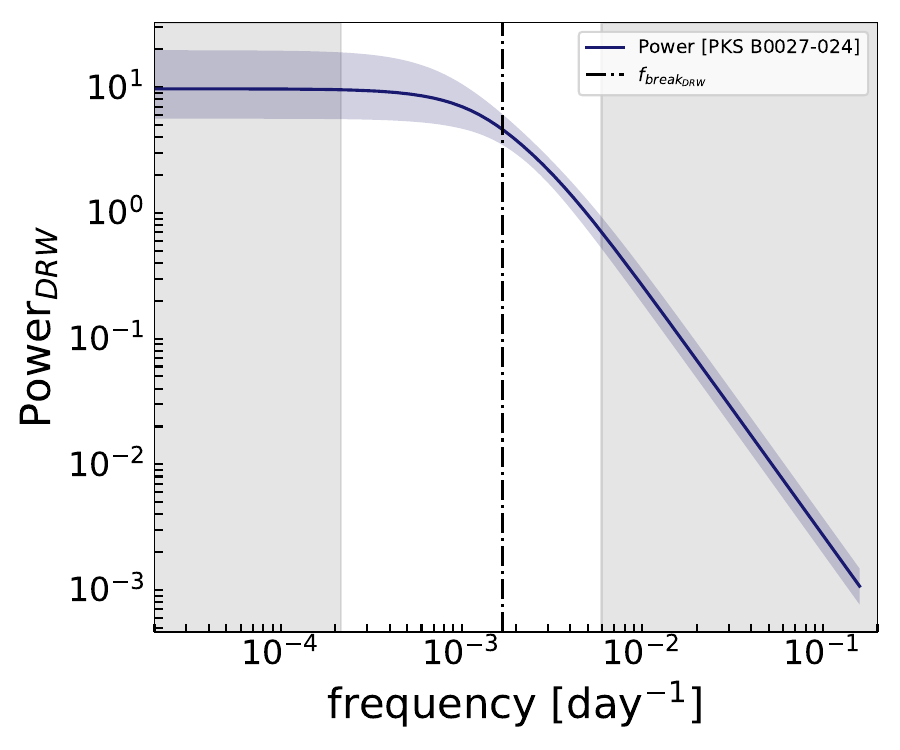}\hspace{1pt}
    \includegraphics[width=0.24\textwidth]{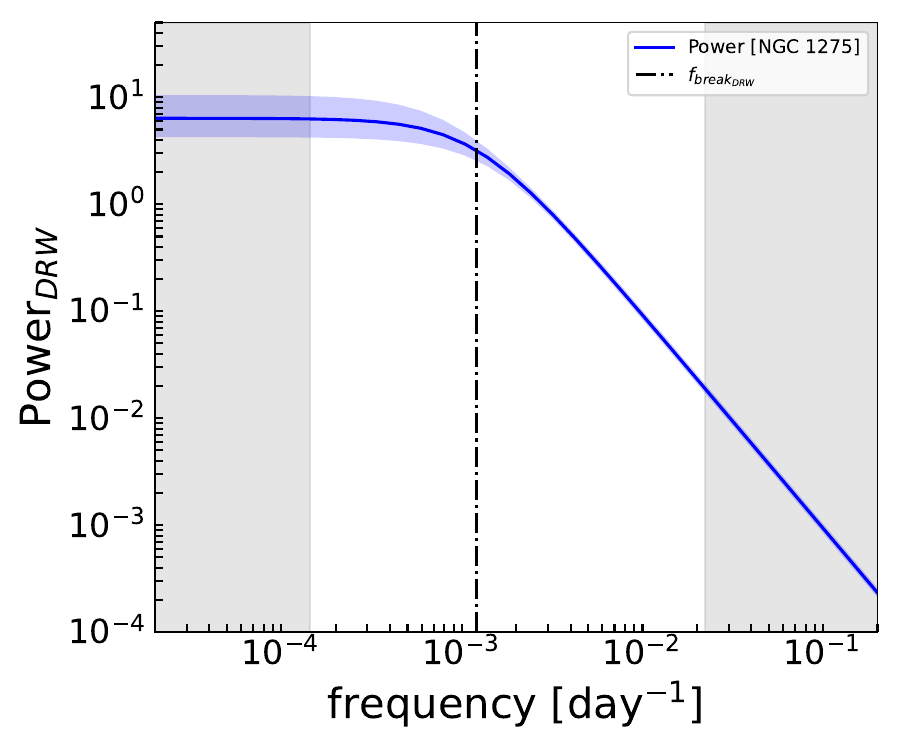}\hspace{1pt}
    \includegraphics[width=0.24\textwidth]{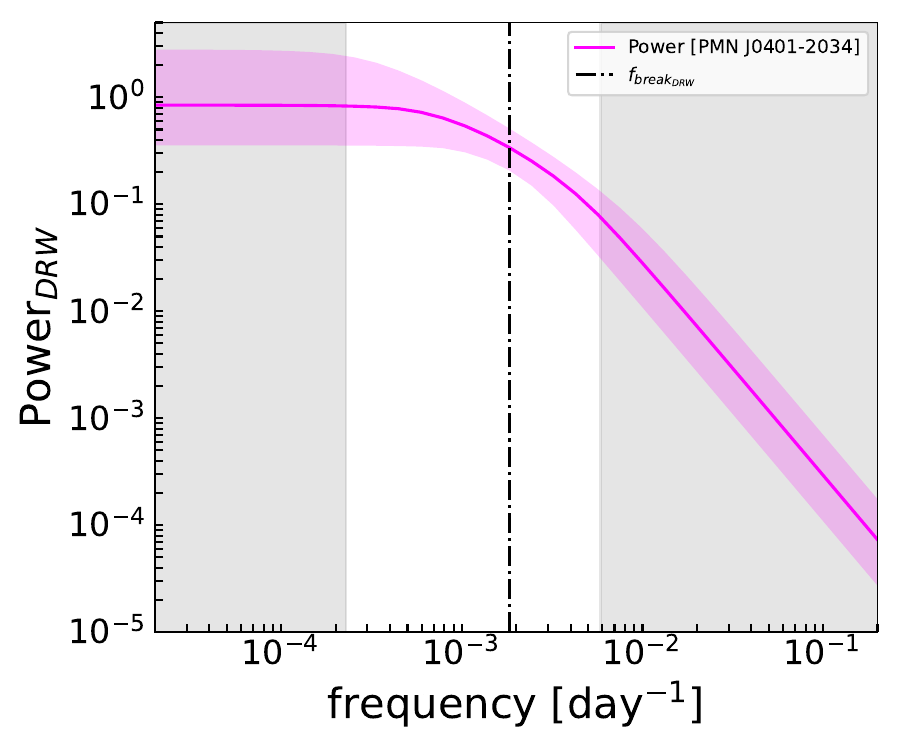}\hspace{1pt}
    \includegraphics[width=0.24\textwidth]{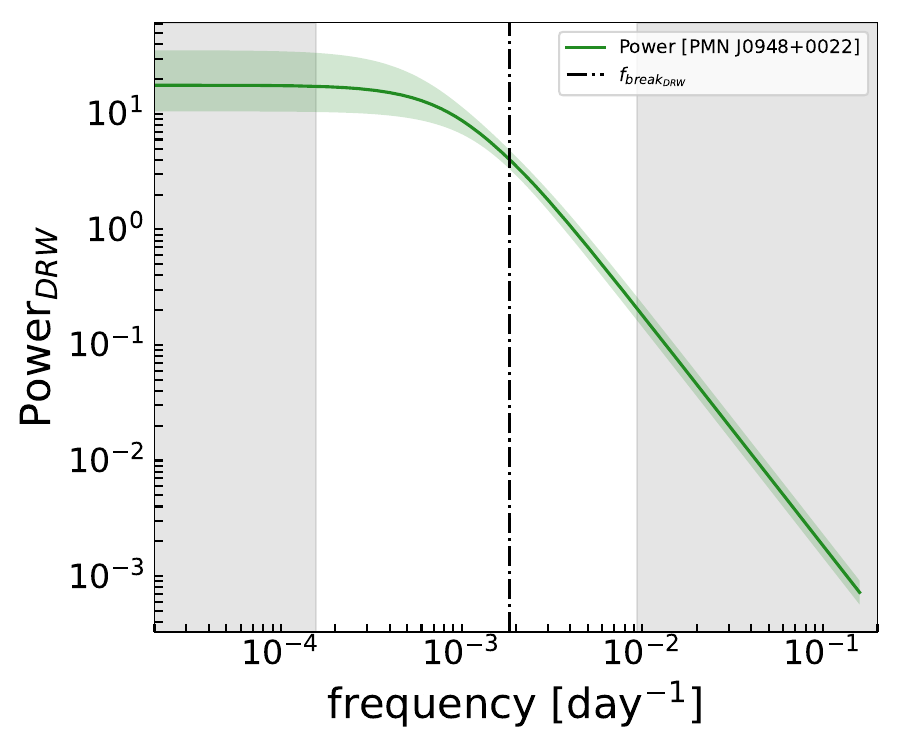}\vspace{1pt}
    \includegraphics[width=0.24\textwidth]{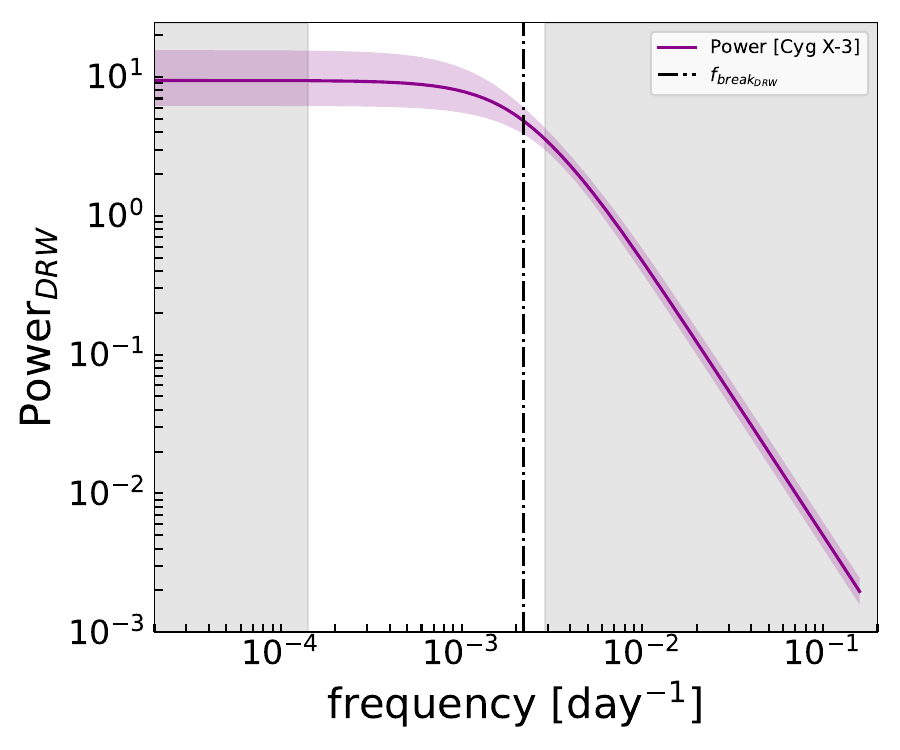}\hspace{1pt}
    \includegraphics[width=0.24\textwidth]{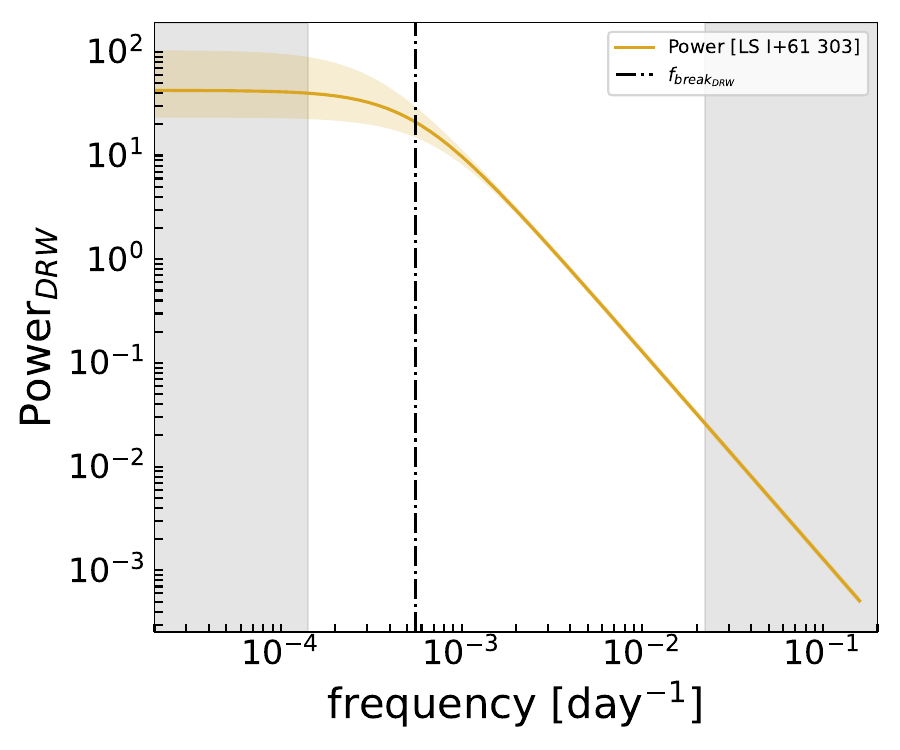}
    \caption{The PSDs of $\gamma$-rays, derived from the DRW modeling results, are presented in this figure. Each PSD profile is shown with a shaded region representing the 1$\sigma$ confidence interval, corresponding to the respective colors. Additionally, two grey-shaded regions in each panel indicate the biased regions caused by the limited length and mean cadence of the light curve. A dotted vertical line marks the break frequency in the PSD.}
    \label{Fig-4}    
\end{figure*}

\begin{figure*}
    \centering
    \includegraphics[width=0.49\textwidth]{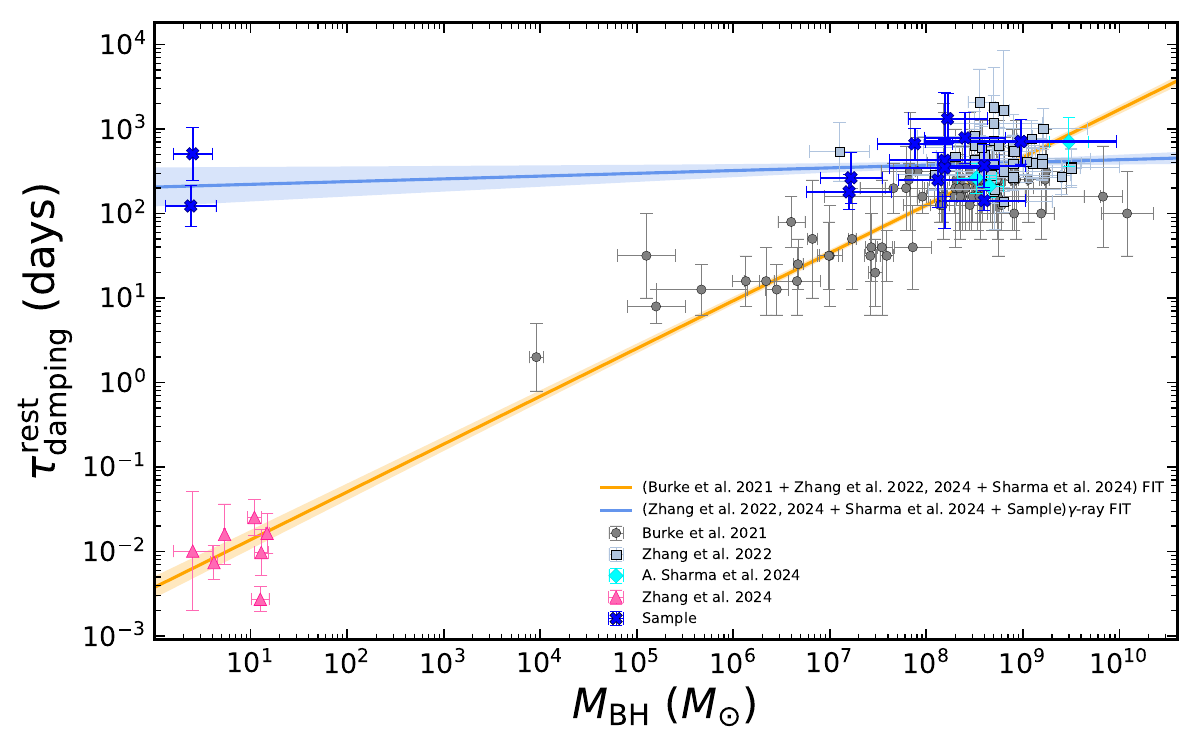}\hspace{1pt}
    \includegraphics[width=0.49\textwidth]{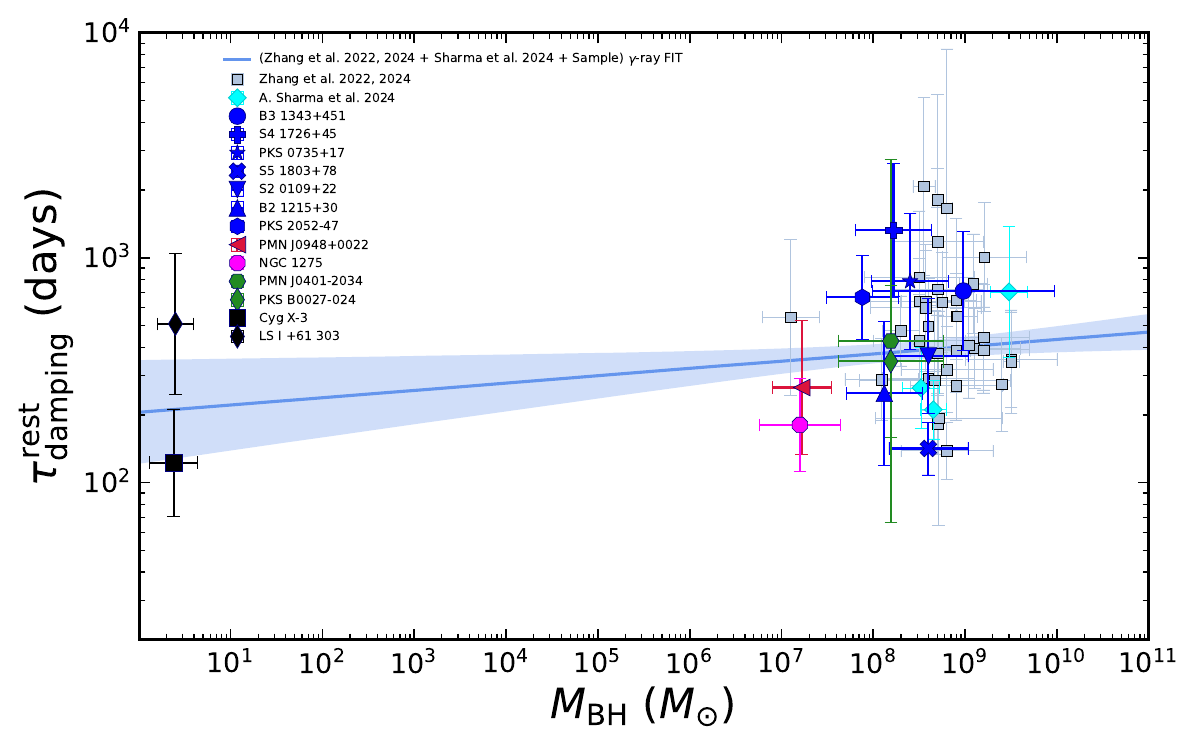}
    \caption{The figure shows the observed $\tau_{DRW}^{rest}$ as a function of $\rm{M_{BH}}$. Data points from different studies are distinguished by various colors and shapes: grey, light blue, cyan, and light pink correspond to \cite{burke2021characteristic, zhang2022characterizing, sharma2024probing, zhang2024discovering}, respectively. In the left panel, an orange line represents a combined fit to the findings from these studies, while the blue line with 1$\sigma$ confidence interval, shows the fit based solely on the $\gamma$-ray results from our sample and  \cite{zhang2022characterizing, sharma2024probing}. The right panel illustrates the same combined fit to the $\gamma$-ray data exclusively, with sources from different classes in our sample indicated by distinct colors and shapes: blazars (blue), NLS1 galaxy (red), radio galaxy (magenta), BCUs (green), and microquasars (black).} 
    \label{Fig-5}    
\end{figure*}


\section{RESULTS}\label{sec:result}

We employed DRW modeling to explore the variability characteristics timescale in the $\gamma$-ray lightcurves of 7 blazars, as listed in Table \ref{tab-1}. The results are illustrated in Figure \ref{Fig-1}. To assess the quality of the DRW modeling, we analyzed the probability densities of the Autocorrelation Functions (ACF) of both the residuals and the squared residuals for all blazars. The lagged values in ACFs consistently fell within the 95$\%$ confidence interval expected for white noise. Additionally, the residuals' distribution was fitted to a normal distribution, with the resulting parameters detailed in Table \ref{tab-1}. The distribution showed good agreement with a normal distribution, characterized by a mean ($\mu$) value close to zero and a standard deviation ($\sigma$) less than one. This finding is further validated by the Kolmogorov–Smirnov (KS) test\footnote{\url{https://docs.scipy.org/doc/scipy/reference/generated/scipy.stats.kstest.html}}, which was applied to evaluate the normality of the residuals. The p-values of the test observed for all blazars are greater than 0.05, indicating that the null hypothesis of normality can not be rejected, see the Table \ref{tab-1}.\par
A similar approach was applied to analyze the light curves of non-blazars and microquasars, and the results are presented in Figure \ref{fig-2} and Table \ref{tab-1}. For a few sources, the observed lagged values in the ACFs fall outside the 95$\%$ confidence interval, suggest that the model may not have adequately captured the flaring episodes in the light curve, or indicate that the events observed in the light curve are stochastic in nature.\par
The posterior probability density distributions of the parameters obtained from the modeling using the \texttt{MCMC} algorithm are displayed in the Figure \ref{Fig-4} and the best-fitting values are listed in Table \ref{tab-1}.\par
Figure \ref{Fig-1}, \ref{fig-2} present these findings in three columns with two rows for each source. The leftmost column displays the $\gamma$-ray lightcurve fitted with the DRW model in the top row, with the corresponding residuals shown in the bottom row. The middle column shows the distribution of the residuals alongside the best-fit normal distribution profile. The rightmost column features the ACFs of the residuals and the squared residuals.\par
However, it's important to consider how the limited length of a light curve can influence the modeling results. Insufficient length can bias the measurement of the damping timescale, leading to potential inaccuracies \cite{kozlowski2017limitations, suberlak2021improving}. To ensure the reliability of the characteristic timescale measurement, we followed the criteria established by \cite{burke2021characteristic}. According to this criteria, the length of the light curve should be at least 10 times the timescale of interest ($\tau_{DRW} <0.1\times$ baseline) and the observed damping timescale should be greater than the average cadence of the light curve ($\tau_{DRW} > \rm{cadence}$). The mean cadence and baseline of each source in our sample is given in the Table \ref{tab-1}. We identified biased regions corresponding to timescales greater than 20$\%$ of the light curve length and less than the mean cadence. As shown in Figure \ref{Fig-4}, the shaded areas in the Power Spectral Density (PSD) indicate unreliable regions where the PSD is not adequately sampled. For all sources in our sample, the observed damping timescale values fall outside these unreliable regions, suggesting that the model effectively captured the characteristics of the light curve.\par
Applying the reliability criteria in the estimation of variability timescale as mentioned above, we obtained the damping timescale for 7 Blazars, 2 BCUs, 1 Radio galaxy, 1 Narrow-line 1 galaxy, and 2 Microquasars. The observed damping timescale ($\tau_{DRW}$) of our sources in observer's frame of reference spans from 40 to $\sim$300 days, which is reliable according to \cite{burke2021characteristic}. However, since the characteristic timescale of variability was obtained in the observer's frame, it must be corrected for the Doppler beaming effect and redshift of each source to determine the timescale in the rest frame. This can be calculated using the following equation:

\begin{equation}
    \tau_{damping}^{rest} = \frac{\tau_{obs}}{1+z} \delta_{D}
\end{equation}

where z is the redshift of the source and $\delta_D$ is the Doppler factor. Estimating the Doppler factor for AGNs is a challenging task, often involving various methods such as modeling the broadband SED \cite{chen2018jet, pei2020estimation, zhang2020doppler} and constraining the equipartition brightness temperature of radio flares \cite{liodakis2017f}. Despite these approaches, determining the Doppler factor remains subject to significant uncertainties. For the sources in our sample, we used the $\delta_D$ values obtained from a literature survey, with references provided in Table \ref{tab-1}. The variability timescale in optical observations from AGN accretion disc has been extensively studied \cite{collier2001characteristic, kelly2009variations, macleod2010modeling, simm2016pan, suberlak2021improving, burke2021characteristic, zhang2023gaussian}. More recently, similar studies have been conducted in the submillimeter range \cite{chen2023testing}, X-rays \cite{zhang2024discovering}, and $\gamma$-rays \cite{ryan2019characteristic, zhang2022characterizing, sharma2024probing, zhang2024discovering}. In this study, we analyzed the $\gamma$-ray light curves of 11 AGNs and 2 microquasars and observed timescale value for microquasars is similar to AGN. When comparing our observed $\gamma$-ray rest-frame timescales, calculated using Equation (1), with the findings from \cite{burke2021characteristic}, \cite{zhang2022characterizing}, \cite{sharma2024probing}, and \cite{zhang2024discovering}, the fit was not very satisfactory. This discrepancy is due to significant variations in the damping timescales across different wavebands for microquasars. Consequently, we fit our variability timescale data exclusively with $\gamma$-ray variability timescales. The best-fit curve, shown as a blue line with a 1$\sigma$ confidence interval, the right panel in Figure \ref{Fig-5}. The best-fitting relation is

\begin{equation}
    \tau_{\mathrm{damping}}^{\mathrm{rest}} = 346.54_{-48.11}^{+41.91} \left( \frac{M}{10^8 M_{\odot}} \right)^{0.06_{-0.03}^{+0.03}}
\end{equation}

with an intrinsic scatter of 0.17$\pm$0.041 and a Pearson correlation coefficient r=0.206, indicating a weak or no correlation between variability timescale in $\gamma$-rays and Black Hole mass, as shown in Figure \ref{Fig-5}. In our investigation, the observed $\tau_{DRW}$ values for NGC 1275 and B2 1215+30 are differ from the findings in the literature \cite{zhang2022characterizing} and \cite{zhang2024discovering}, respectively. This discrepancy may be due to the use of longer light curves.



\section{DISCUSSION}\label{sec:discussion}
This work is a follow-up to our previous study \cite{sharma2024probing}, where we modeled the $\gamma$-ray light curves of three blazars with the DRW model and compared the results with findings from the literature. That study inspired us to further investigate the variability characteristics of different classes of AGNs and expand it to include the stellar black hole mass systems. To achieve this, we utilized over a decade of $\gamma$-ray light curve data from the Fermi-LAT observatory. While Fermi-LAT detections are predominantly blazars, it has also successfully captured several non-blazar sources. By analyzing these diverse sources, we aim to understand the emission processes responsible for the observed variations in their light curves.\par

In this work, We present a comprehensive investigation focused on characterizing the $\gamma$-ray variability of 7 blazars, 1 radio galaxy, 1 narrow-line Seyfert 1 galaxy, and 2 unclassified blazar candidates, and 2 microquasars. Such studies offer valuable insights into the high-energy emission regions within relativistic jets and the inner jet structure. It is widely believed that black holes, from stellar mass to supermassive, are often associated with relativistic jets. This study provides indirect evidence supporting a universal scaling law, suggesting that the jet production mechanism is invariant of their black hole mass.\par
The Gaussian Process (GP) method has been extensively used to characterize variability across different wavebands in a wide range of astrophysical objects, from low-mass microquasars to AGNs with supermassive black holes. It serves as a powerful tool for studying the statistical properties of variability in these objects. Several studies has been carried out to explore the characteristic timescale of optical variability \cite{collier2001characteristic, kelly2009variations, macleod2010modeling, simm2016pan, goyal2018stochastic, ryan2019characteristic, covino2020looking, tarnopolski2020comprehensive, suberlak2021improving,burke2021characteristic, yang2021gaussian,zhang2023gaussian}. \cite{ruan2012characterizing} modeled optical light curves from December 2002 to March 2008 and observed variability timescales longer than those found in Fermi-detected blazars. \cite{xiong2015basic} identified significant differences in physical properties such as redshift, black hole mass, jet kinetic power, and broad-line luminosity between Fermi-detected and non-Fermi-detected blazars. \cite{paliya2017general} reported that non-Fermi-detected blazars tend to have smaller Doppler factors. \cite{burke2021characteristic} suggested that the DRW damping timescale, measured from accretion disc variability in normal quasars, may correspond to the thermal instability timescale predicted by AGN standard accretion disc theory.\par
\cite{chen2023testing} explored accretion physics by investigating submillimeter variability in low-luminosity AGNs, finding that the submillimeter emission has a significantly shorter variability timescale compared to optical emission, likely linked to the innermost regions of the black hole system. They also examined X-ray variability and found it to exhibit similarly short timescales, indicating that similar processes may be driving both emissions.\par
Further, \cite{zhang2024discovering} extended similar kind of study to microquasars with X-rays, observing a scaling relationship, $\tau \propto M_{BH}$, consistent with \cite{burke2021characteristic}, suggesting that the accretion disc's variability may imprint on jet emission or be associated with physical processes occurring within the jet.\par
\cite{ryan2019characteristic} analyzed Fermi-LAT observations of 13 blazars over $\sim$9.5 years, \cite{zhang2022characterizing} examined a sample of 23 AGNs, including 22 blazars and 1 radio galaxy, with Fermi-LAT data spanning approximately $\sim$12.7 years, and \cite{sharma2024probing} used $\sim$15-year-long $\gamma$-ray light curves of 3 blazars, all modeled using stochastic processes. These samples cover a black hole mass range from $\sim 10^7$ to $10^{10} \rm{M_{\odot}}$. The variability timescales obtained from these models align with the thermal timescales on the $\tau_{DRW}^{rest} - \rm{M_{BH}}$ plane, as described by \cite{burke2021characteristic}, suggesting a strong association with accretion disk variations. This connection can be explained by propagating fluctuation models \cite{lyubarskii1997flicker}, where the jet launching site is in close proximity to the accretion disk, leading to the imprinting of accretion disk variations on jet emissions \cite{kataoka2001characteristic}. Several accretion disk models, such as those proposed by \cite{shakura1973black, siemiginowska1996evolution, kataoka2001characteristic, hirose2009radiation}, are likely candidates for explaining such variability features. If the variability originates in the relativistic jet, the observed timescale would be influenced by the jet's Doppler factor, which is often uncertain \cite{liodakis2015population}. Variations in jet emission could be driven by fluctuations in magnetic field strength, electron injection rate, or Doppler factor \cite{mastichiadis1997variability}.\par
The derived damping timescales from the $\gamma$-ray emissions of all sources in our sample are plotted against $\rm{M_{BH}}$ in Figure \ref{Fig-5}. In this figure, the timescales for high-mass AGNs overlap with the thermal timescales from \cite{burke2021characteristic} and the non-thermal timescales from \cite{zhang2022characterizing} and \cite{sharma2024probing}. Additionally, the X-ray variability timescales from microquasars, as reported by \cite{zhang2024discovering}, are also included in the $\tau_{DRW}^{rest} - \rm{M_{BH}}$ plot. These X-ray timescales exhibit a relationship with both thermal and non-thermal timescales from earlier studies and our own findings in high-mass AGNs. As discussed in Section \ref{sec:result}, the observed $\gamma$-ray variability timescales in both microquasars are comparable in magnitude to those in high-mass AGNs and are significantly larger than the timescales observed in other wavebands.\par 

We first compared the characteristic time scale among various types of AGN and it was found that the luminous FSRQ, NLSy1, radio galaxy, and unknown blazar type also show a similar variability time scale, referring to the fact that jet properties remain the same across the AGN types. The jet luminosity among these objects can differ however, the variability pattern remains universal. 

We also included stellar mass black holes and estimated their timescales in the $\gamma$-ray emission of microquasars Cyg X-3 and LS I+61 303 that were not detected earlier. The rest-frame timescales for both the objected are approximately 186 days and 446 days, respectively. These timescales are comparable to those observed in AGNs. Figure \ref{Fig-5} shows a best-fitting correlation between the $\gamma$-ray timescales of low-mass microquasars and high-mass AGNs. This raises the question of whether the emission from relativistic jets depends on the mass of the host black hole. Are the emission mechanisms within the jets or the processes involved in jet production independent of black hole mass? These are critical questions that need to be addressed.\par

\cite{liodakis2017scale} found a strong correlation between the intrinsic broadband radio luminosity of blazars and black hole mass, extending this relationship to the stellar black hole mass systems like microquasars. This finding imposes significant constraints on alternative jet models. This study suggests that the jet production mechanims is a universal process common in all black hole systems independent to their black hole mass and AGNs also operate in a similar accretion regime as the hard state in microquasars. This proposed methodology also stand out as an independent method for estimating the Doppler factor.\par
The findings from this investigation also suggest indirect evidence that the variability observed in $\gamma$-rays, driven by physical processes within relativistic jets, may be universal across black hole systems, regardless of their mass. While jets in microquasars extend over shorter distances compared to those in AGNs, they are still efficient enough to produce similar emission variations. It is also possible that the observed variability in both microquasars and AGNs arise from distinct processes occurring in different emission regions or within the jets themselves.


\section{CONCLUSION}\label{conclusion}
We present the $\gamma$-ray variability of 11 AGNs (FSRQ, BCUs, NLSy1, and radio galaxy) and 2 microquasars utilizing $\sim$15 yr of the Fermi-LAT light curve. To characterize this variability, we modeled the light curves with a DRW model. Our main findings are pointed out below.
\begin{itemize}
\item We noted that even though FSRQs, BCUs, NLSy1, and radio galaxies are of different kinds. but the jets they host have similar properties.
\item Our findings indicate that the observed characteristic variability timescales for both AGNs and microquasars are remarkably similar, irrespective of their black hole masses.
\item This suggests that common underlying processes may be responsible for the observed variations in $\gamma$-ray emissions across these sources.
\item The same variability time scale in stellar Black hole and supermassive black hole suggests the jet possibly can be produced by a similar procedure. 
\end{itemize}






\section*{Acknowledgements}
We acknowledge the various software used for the analysis and the Fermi-LAT archival data center.


\bibliographystyle{elsarticle-harv} 
\bibliography{example}

\appendix

\section{Appendix title 1}
Table \ref{tab-1} contains information of all sources in our sample.

\begin{table*}[htbp] 
\footnotesize
\centering
\setlength{\extrarowheight}{10pt}
\setlength{\tabcolsep}{2pt}

\rotatebox{90}{
\begin{minipage}{\textheight}
\centering
\begin{tabular}{ccccccccccccccccccc}
\hline
\hline
\multirow{2}{*}{Source} & \multirow{2}{*}{4FGL name} & \multirow{2}{*}{R.A.} & \multirow{2}{*}{Dec.} & \multirow{2}{*}{Source} & \multirow{2}{*}{$\left<\rm{cadence}\right>$} & \multirow{2}{*}{Baseline} &\multirow{2}{*}{\textit{z}} & \multirow{2}{*}{$\delta_D$} & \multirow{2}{*}{log} & \multirow{2}{*}{ln$\sigma_{DRW}$} & \multirow{2}{*}{ln$\tau_{DRW}$} & \multirow{2}{*}{log$_{10} \tau_{rest}$} & \multicolumn{3}{c}{Normal Distribution Fitting} & \multicolumn{3}{c}{Ref.}\\
\cline{14-16}
    &   &   &   & type & (days)  & (days)  &   &   & ($M/M_{\odot}$)  &   & (days) & (days) & $\mu$ & $\sigma$ & KS-test & z &$\delta_D$ & $M_{BH}$ \\
    &   &   &   &   &   &   &   &   &   &   &   &   &   &   & (p-value) &  &  \\
(1) & (2) & (3) & (4) & (5) & (6) & (7) & (8) & (9) & (10) & (11) & (12) & (13) & (14) & (15) & (16) & (17) & (18) & (19)\\
\hline
S2 0109+22 & J0112.1+2245 & 18.0243 & 22.7441 & BLL  & 8.81 & $\sim$5642  & 0.36 & 2.59 & 8.6$\pm$0.43 &-0.38$_{-0.12}^{+0.15}$  &5.26$_{-0.26}^{+0.33}$&2.67$_{-0.11}^{+0.14}$&0.032$\pm$0.027&0.26$\pm$0.022&1.0 & (1) & (2) & (3)\\
PKS 0735+17 & J0738.1+1742 & 114.531 & 17.7053 & BLL  & 10.58  & $\sim$5621  & 0.424 & 4.5 & 8.4$\pm$0.42 &-0.42$_{-0.14}^{+0.19}$&5.52$_{-0.30}^{+0.40}$&3.39$_{-0.13}^{+0.18}$&0.033$\pm$0.019&0.21$\pm$0.015&0.98 & (4) & (5) & (4)\\
B2 1215+30 & J1217.9+3007 & 184.467 & 30.1168 & BLL  & 8.47 & $\sim$5628  & 0.13  & 1.1 & 8.12$\pm$0.41 &-0.54$_{-0.14}^{+0.19}$&5.55$_{-0.32}^{+0.42}$&3.54$_{-0.15}^{+0.18}$&0.0422$\pm$0.025&0.311$\pm$0.020&0.92 & (6) & (5) & (7)  \\
B3 1343+451 & J1345.5+4453 & 206.388 & 44.8832 & FSRQ  & 13.59  & $\sim$5586  & 2.534 & 12.6 & 8.98$\pm$0.99 &-0.32$_{-0.12}^{+0.16}$&5.30$_{-0.27}^{+0.34}$&2.85$_{-0.12}^{+0.15}$&-0.0002$\pm$0.007&0.09$\pm$0.006&0.989 & (8) & (9) & (10) \\
S4 1726+45 & J1727.4+4530 & 261.865 & 45.511 & FSRQ & 20.81 & $\sim$5621 & 0.717 & 13.4 & 8.22$\pm$0.411 &-0.42$_{-0.12}^{+0.15}$&5.14$_{-0.31}^{+0.37}$&3.11$_{-0.13}^{+0.16}$&-0.006$\pm$0.009&0.144$\pm$0.007&0.66 & (11) & (9) & (12) \\
S5 1803+78 & J1800.6+7828 & 270.19 & 78.4678 & BLL & 8.40 & $\sim$5642 & 0.68 & 5.97 & 8.6$\pm$0.43 &-0.04$_{-0.06}^{+0.06}$&3.69$_{-0.13}^{+0.14}$&2.53$_{-0.08}^{+0.10}$&-0.008$\pm$0.0007&0.033$\pm$0.0005&0.167 & (13) & (13) & (14)\\
PKS 2052-47 & J2056.2-4714 & 314.068 & -47.2466 & FSRQ & 11.83 & $\sim$5481 & 1.49 & 17.1 & 7.88$\pm$0.39 &-0.29$_{-0.09}^{+0.11}$&4.58$_{-0.20}^{+0.23}$&2.83$_{-0.09}^{+0.10}$&-0.011$\pm$0.007&0.10$\pm$0.006&0.299 & (15) & (15) & (16)  \\
PKS B0027-024 & J0030.6-0212 & 7.6326 & -2.1989 & BCU & 26.78 & $\sim$3696 & 1.804 & 10.3 & 8.19$\pm$0.57 &-0.45$_{-0.13}^{+0.15}$&4.55$_{-0.36}^{+0.42}$&2.64$_{-0.16}^{+0.18}$&-0.006$\pm$0.020&0.168$\pm$0.017&0.786 & (17) & (9) & (18) \\
NGC 1275 & J0319.8+4130 & 49.9507 & 41.5117 & RG & 7.20 & $\sim$5495 & 0.0176 & 1.4 & 7.2$\pm$0.44 &-0.48$_{-0.10}^{+0.12}$&4.88$_{-0.22}^{+0.26}$&2.25$_{-0.09}^{+0.12}$&0.004$\pm$0.003&0.108$\pm$0.002&0.98 & (20) & (20) & (20)  \\
PMN J0401-2034 & J0401.9-2034 & 60.4696 & -20.5861 & BCU & 27.17 & $\sim$3479 & 1.647 & 13.1 & 8.19$\pm$0.57 &-1.26$_{-0.18}^{+0.19}$&4.46$_{-0.82}^{+1.04}$&2.62$_{-0.36}^{+0.45}$&-0.042$\pm$0.024&0.215$\pm$0.02&0.73 & (19) & (9) & (18)  \\
PMN J0948+0022 & J0948.9+0022 & 147.239 & 0.3737 & NLS1 & 17.11 & $\sim$5042 & 0.5846 & 2.7 & 7.22$\pm$0.32 &-0.40$_{-0.12}^{+0.15}$&5.05$_{-0.31}^{+0.38}$&2.50$_{-0.13}^{+0.17}$&-0.025$\pm$0.009&0.116$\pm$0.007&0.786 & (21) & (27) & (21) \\
Cyg X-3 & J2032.6+4053 & 308.107 & 40.9577 & Microquasar & 55.31  & $\sim$5642  & 0.003 & 1.78 & 0.38$\pm$0.26 &-0.31$_{-0.10}^{+0.11}$&4.24$_{-0.26}^{+0.29}$&1.83$_{-0.11}^{+0.12}$&0.003$\pm$0.005&0.035$\pm$0.004&0.351 & (22) & (25) & (23) \\
LS I+61 303 & J0240.5+6113 & 40.1319 & 61.2293 & Microquasar & 7.13  & $\sim$5620  & 0.0013 & 1.78 & 0.6$\pm$0.2 &-0.28$_{-0.15}^{+0.21}$&5.66$_{-0.30}^{+0.42}$&2.45$_{-0.13}^{+0.18}$&-0.001$\pm$0.007&0.263$\pm$0.006&0.999 & (24) & (25) & (26) \\
\hline
\end{tabular}

\vspace{5pt} 

\caption{\label{tab: qpo-timescale} This table provides detailed information on 12 Active Galactic Nuclei (AGNs)—including blazars (BL Lac or FSRQ), radio galaxy (RG), narrow-line Seyfert 1 galaxies (NLS1), and blazar candidates of unclassified type (BCU)—and 2 microquasars. The columns are defined as follows: (1) Source Name, (2) 4FGL Name, (3)-(4) Coordinate of the source, (5) Source Type, (6) Mean cadence (days), (7) Duration of light curve (days) (8) Redshift (z), (9) Doppler Factor ($\delta_D$), (10) Black Hole Mass (in $M_{\odot}$ unit), (11)-(12) Posterior parameters of modeling the Lightcurves with the DRW process, (13) Damping timescale in the rest frame (days), (14)-(15) Parameters of the Normal distribution fitted to the residuals, (16) Kolmogorov–Smirnov (KS) Test result, (17)-(19)References for the redshift (z), Doppler factor ($\delta_D$), and black hole mass values:(1)\cite{magic2018broad}, (2)\cite{fan2013gamma}, (3)\cite{zhang2023detection}(4)\cite{chai2012governs}, (5)\cite{liodakis2017f}, (6)\cite{furniss2019spectroscopic}, (7)\cite{valverde2020decade}, (8)\cite{sahakyan2020exploring}, (9)\cite{chen2018jet}, (10)\cite{zhang2024fundamental}, (11)\cite{costamante2018origin}, (12)\cite{xiong2014intrinsic}, (13)\cite{ghisellini2010general}, (14)\cite{lin2017intrinsic}, (15)\cite{wang2022possible}, (16)\cite{kadowaki2015role}, (17)\cite{pena2021optical}, (18)\cite{xiao2022relativistic}, (19)\url{https://ned.ipac.caltech.edu}, (20)\cite{zhang2022characterizing}, (21)\cite{xin2022multicolor}, (22)\cite{koljonen2020obscured}, (23)\cite{zdziarski2013cyg}, (24)\url{https://simbad.u-strasbg.fr/simbad/sim-fbasic}, (25)\cite{molina2019model}, (26)\cite{massi2017black}, (27)\cite{doi2019radio}}
\end{minipage}}
\label{tab-1}
\end{table*}








\end{document}